\documentclass[10pt]{iopart}

\usepackage{iopams}
\usepackage{graphicx}
\usepackage{xcolor}

\begin{document}

\title{Peculiar Phase Diagram with Isolated Superconducting Regions in ThFeAsN$_{1-x}$O$_x$}

\author{Bai-Zhuo Li,$^{1}$ Zhi-Cheng Wang,$^{2}$ Jia-Lu Wang,$^{4}$ Fu-Xiang Zhang,$^{1}$ Dong-Ze Wang,$^{1}$ Feng-Yuan Zhang,$^{1}$ Yu-Ping Sun,$^{1}$ Qiang Jing,$^{1}$ Hua-Fu Zhang,$^{1}$ Shu-Gang Tan,$^{1}$ Yu-Ke Li,$^{4}$ Chun-Mu Feng,$^{2}$ Yu-Xue Mei,$^{1\dag}$ Cao Wang,$^{1\ddag}$ Guang-Han Cao$^{2,3}$}

\address{$^{1}$Department of Physics, School of Physics $\&$ Optoelectronic Engineering, Shandong University of Technology, Zibo 255000, P. R. China\\
         $^{2}$Department of Physics, Zhejiang University, Hangzhou 310027, P. R. China\\
         $^{3}$Collaborative Innovation Centre of Advanced Microstructures, Nanjing 210093, P. R. China\\
         $^{4}$Department of Physics, Hangzhou Normal University, Hangzhou 310036, P. R. China }
\ead{\dag yuxue\_mei@126.com}
\ead{\ddag wangcao@sdut.edu.cn}

\vspace{12pt}
\begin{indented}
\item[]May 2018
\end{indented}

\begin{abstract}

ThFeAsN$_{1-x}$O$_x$ ($0\leq x\leq0.6$) system with heavy electron doping has been studied by the measurements of X-ray diffraction, electrical resistivity, magnetic susceptibility and specific heat. The non-doped compound exhibits superconductivity at $T_\mathrm{c}^\mathrm{onset}=30$ K, which is possibly due to an internal uniaxial chemical pressure that is manifested by the extremely small value of As height with respect to the Fe plane. With the oxygen substitution, the $T_\mathrm{c}$ value decreases rapidly to below 2 K for $0.1\leq x\leq0.2$, and surprisingly, superconductivity re-appears in the range of $0.25\leq x\leq0.5$ with a maximum $T_\mathrm{c}^\mathrm{onset}$ of 17.5 K at $x=0.3$. For the normal-state resistivity, while the samples in intermediate non-superconducting interval exhibit Fermi liquid behavior, those in other regions show a non-Fermi-liquid behavior. The specific heat jump for the superconducting sample of $x=0.4$ is $\Delta C/(\gamma T_\mathrm{c})=0.89$, which is discussed in terms of anisotropic superconducting gap. The peculiar phase diagram in ThFeAsN$_{1-x}$O$_x$ presents additional ingredients for understanding the superconducting mechanism in iron-based superconductors.

\end{abstract}

%
%
%
%
%

\section{Introduction}

Discovered in the year 2008, superconductivity in a group of iron-based compounds becomes one of the hottest topics in condensed matter physics.\cite{Hosono LaFeAsO-F} It was not long before scientists realized that doping with either electrons or holes may induce superconductivity.\cite{Hosono LaFeAsO-F,LaFeAsO-Sr,Ba122-K} On this basis, the substitution phase diagrams about tens of individual dopants have been established.\cite{Johnston review, ChenXH review} In the prototype ``1111" system, the carrier doping level is mostly limited by the heterovalent substitution solubility, such that the overdoped regime cannot be reached.\cite{Hosono LaFeAsO-F,ChenXH SmFeAsO-F,cxh,GdFeAsO-Th} However, recent experimental works show that the high-pressure synthesis technique can remarkably increase the substitution solubility and, the electron doping level is pushed to 0.53 electrons/Fe-atom in  LaFeAsO$_{1-x}$H$_x$,\cite{LaFeAsO-H two dome} and 0.75 electrons/Fe-atom in LaFeAsO$_{1-x}$F$_x$.\cite{LaFeAsO-F two-dome} A secondary superconducting dome with a higher maximum superconducting transition temperature ($T_\mathrm{c}$) was observed in both systems. These findings shed light on the superconducting mechanisms in iron-based superconductors.\cite{two dome theory}

 In a previous work of our group, we reported the discovery of a new ``1111'' type iron-based compound ThFeAsN, which superconducts below 30 K without external chemical doping.\cite{ThFeAsN} Though theoretical calculations suggest a striped anti-ferromagnetic (AFM) ground state,\cite{ThFeAsN-band1, ThFeAsN-band2} the studies on $^{57}$Fe M\"{o}ssbauer spectroscopy, neutron powder diffraction, and $\mu$SR/NMR experiments found no magnetic order of the Fe moments down to 2.0 K.\cite{ThFeAsN-Mossbauer,ThFeAsN-neutron,ThFeAsN-NMR, ThFeAsN-NMR-2} To understand the absence of AFM ordering and the emergence of superconductivity in non-doped ThFeAsN, it is vitally necessary to look at the evolution of the superconducting phase with (preferably heavy) electron doping. In this work, we study the oxygen substitution effect in ThFeAsN$_{1-x}$O$_x$. Unexpectedly, we find that the \emph{nominal} oxygen solubility reaches as high as $x=0.6$ under ambient pressure. The measurement of electronic resistivity, as well as magnetic susceptibility, indicates that the 30 K superconductivity in ThFeAsN is rapidly suppressed to below 2 K at $x=0.1$. Following the quenching of superconductivity within $0.1\leq x\leq0.2$, superconductivity re-appears in the region $0.25\leq x\leq0.5$, with a maximum $T_\mathrm{\mathrm{c}}$ of 17.5 K. While the two-superconductivity-region phenomenon somewhat resembles the aforementioned LaFeAsO$_{1-x}$H$_x$ and LaFeAsO$_{1-x}$F$_x$ systems,\cite{LaFeAsO-H two dome,LaFeAsO-F two-dome} the present system is remarkable for the isolated superconducting windows and, for the lower maximum $T_\mathrm{c}$ in the second superconducting region.

\section{Experimental details}

Polycrystalline samples of ThFeAsN$_{1-x}$O$_{x}$ ($0\leq x\leq0.6$) were synthesized using powder of Th$_3$N$_4$, ThO$_2$, Th and FeAs as starting materials. The preparation of the precursors and the sintering condition of the final products are similar to those of the ThFeAsN parent compound.\cite{ThFeAsN} Powder x-ray diffraction (XRD) was carried out at room temperature on a PANalytical X-ray diffractometer (Model EMPYREAN) with a monochromatic Cu K$\alpha1$ radiation. Crystal structure data were obtained by Rietveld refinement using the step-scan XRD data with $20^\circ\leq2\theta\leq 120^\circ$ for all the samples. During the structural refinements, we fix the oxygen content at the nominal value as it is not reliable to detect light elements using the XRD technique. The typical $R$-factors of the refinements are: $R_\mathrm{F}\approx3\%$, $R_\mathrm{B}\approx4\%$, and $R_\mathrm{wp}\approx5\%$, indicating the good reliability of the refinement.\cite{Rietveld} Magnetic measurements were performed on a Quantum Design Magnetic Property Measurement System (MPMS-XL5). The temperature-dependent resistivity was measured using a standard four-terminal method on a Cryogenic Mini-CFM measurement system equipped with a Keithley 2400 digital sourcemeter and a Keithley 2182 nanovoltmeter.

\section{Results and discussion}
\subsection{Crystal structure}

\begin{figure}
\includegraphics[width=7.5cm]{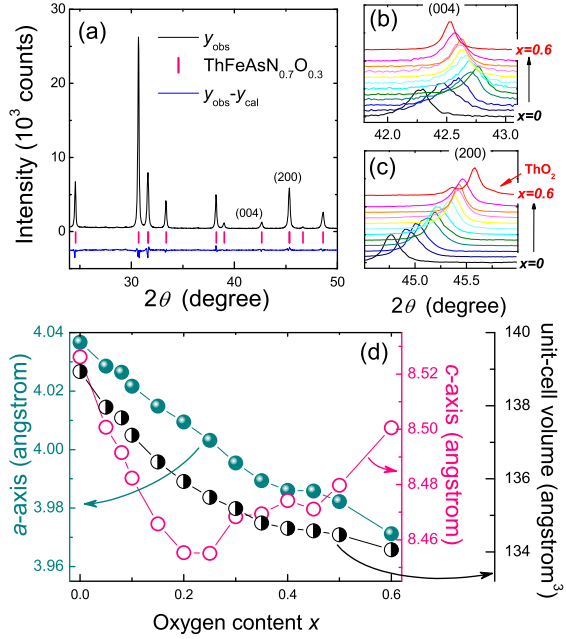}
\caption {(Color online) (a) Rietveld refinement profile (in part) of ThFeAsN$_{0.7}$O$_{0.3}$. (b) and (c) Magnified XRD patterns indicating the peak shift with oxygen substitutions. (d) Lattice parameters and unit-cell volume as functions of the nominal oxygen content in ThFeAsN$_{1-x}$O$_x$. }
\label{fig1}
\end{figure}

Figure \ref{fig1}(a) shows the XRD pattern of the synthesized ThFeAsN$_{0.7}$O$_{0.3}$ sample. The XRD peaks can be well indexed with a tetragonal unit cell of $a$ = 3.9955(4) {\AA} and $c$ = 8.4683(7) {\AA}. No obvious impurity peak is found for the samples with $x\leq0.5$, suggesting that oxygen is successfully incorporated into the lattice (the systematic changes in lattice parameters shown in Figure \ref{fig1}(d) confirm this point). For the samples of $x=0.6$ and 0.7, the impurity peaks begin to show up in the XRD pattern, indicating that the oxygen solubility limit is near $x=0.6$. Note that, such a high solubility can be achieved only by means of high-pressure synthesis in $Ln$FeAsO$_{1-x}$H$_x$ ($Ln$ stands for lanthanides) and LaFeAsO$_{1-x}$F$_x$ systems.\cite{LaFeAsO-H two dome,LaFeAsO-F two-dome} Thus the ThFeAsN$_{1-x}$O$_x$ system is quite unique for the capacity of heavy electron doping under ambient pressure.

Figure \ref{fig1}(b) and (c) show the doping-dependent shift of the two separate reflections, (200) and (004), which are directly related to the $a$ and $c$ axes, respectively. While the (200) reflection drifts towards higher angles upon oxygen doping, the (004) peak shifts to higher angles for $x\leq0.2$, and then it moves to the opposite direction for $x\geq0.25$. Figure \ref{fig1}(d) plots the lattice parameters, obtained from the Rietveld analyses, as functions of nominal oxygen content. Indeed, the $a$-axis steadily shrinks with the oxygen substitution. In contrast, the $c$-axis first goes down rapidly to 8.4551(7) {\AA} at $x=0.25$, and then gradually increases to 8.5006(2) {\AA} at $x=0.6$. To our knowledge, such a non-monotonic change in $c$-axis has not been seen in other iron-based superconductors where the electron doping always leads to the monotonic decrease in $c$.\cite{Hosono LaFeAsO-F,ChenXH SmFeAsO-F,GdFeAsO-Th,LaFeAsO-H two dome,BaFe2As2-Ni}

\begin{table}
\caption{\label{tab.1}Crystallographic data of ThFeAsN$_{1-x}$O$_x$ ($x$=0.3 and 0.6) at room temperature. The space group is P$4/nmm$. The atomic
coordinates are as follows: Th (0.25, 0.25, $z$); Fe (0.75, 0.25, 0.5); As (0.25, 0.25, $z$); N/O (0.75, 0.25, 0). $H_\mathrm{Th-N/O}$, $H_\mathrm{Th-As}$ and $H_\mathrm{Fe-As}$ represent the distance along the $c$-axis between Th and N/O, Th and As, Fe and As, respectively.}
\begin{indented}
\item[]\begin{tabular}{ccc}
\hline
Compounds&ThFeAsN$_{0.7}$O$_{0.3}$&ThFeAsN$_{0.4}$O$_{0.6}$\\
\hline
$a$ ({\AA}) & 3.9955(4) & 3.97125(9)\\
$c$ ({\AA}) & 8.4683(7) & 8.5006(2)\\
$R_\mathrm{wp}$ (\%) &4.37 & 5.21\\
$R_\mathrm{exp}$ (\%) & 3.73 & 3.55\\
$V$ ({\AA}$^3$) & 135.19(2) &  134.060(5)\\
$z$ of Th & 0.14750(7)& 0.1523(1)\\
$z$ of As & 0.6606(2) & 0.6626(2)\\
$H_\mathrm{Th-N/O}$ ({\AA}) & 1.2491(7) &1.2794(9)\\
$H_\mathrm{Th-As}$ ({\AA}) & 1.626(2) &1.591(3)\\
$H_\mathrm{Fe-As}$ ({\AA}) & 1.360(2) &1.369(2)\\
As-Fe-As angle ($^{\circ}$) & 111.5(1) &110.3(1)\\
\hline
\end{tabular}
\end{indented}
\end{table}

\begin{figure}
\includegraphics[width=7.5cm]{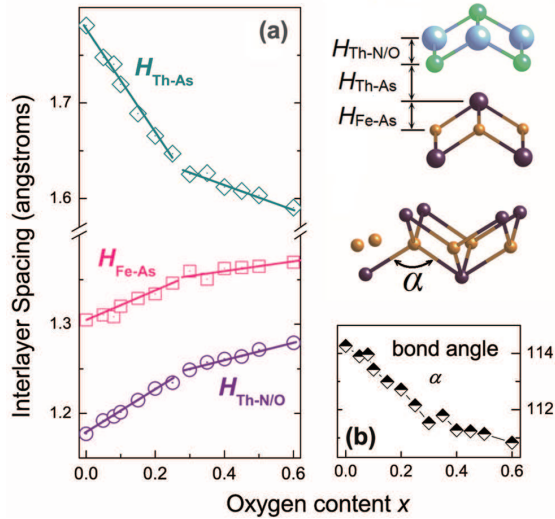}
\caption {(Color online) (a) Interplanar spacings, $H_{\rm{Th-As}}$, $H_{\rm{Fe-As}}$, and $H_{\rm{Th-N/O}}$ (see the upper right structure), as functions of nominal oxygen content in ThFeAsN$_{1-x}$O$_x$. The solid lines are guides to the eye. (b) Oxygen-content dependent diagonal As$-$Fe$-$As bond angle.}
\label{fig2}
\end{figure}

We plot selected crystallographic parameters versus nominal oxygen content in Figure \ref{fig2}. Generally speaking, the substitution of O$^{2-}$ for N$^{3-}$ introduces extra positive (negative) charges in [Th$-$N/O] ([Fe$-$As]) block layers, that enhances the interlayer Coulomb attraction. As shown in Figure \ref{fig2}(a), the spacing between Th and As planar layers ($H_{\rm{Th-As}}$) decreases monotonically from 1.781 {\AA} to 1.591 {\AA}. In contrast, the distance between As and Fe planar layers ($H_{\rm{Fe-As}}$) and the one between Th and N/O planar layers ($H_{\rm{Th-N/O}}$) abnormally increase with oxygen doping. We note that all the three data lines in Figure \ref{fig2}(a) show a kink at $x\sim0.25$, coincident with the minimum of $c$-axis in Figure \ref{fig1}(d). For $x\geq0.25$, all the interplanar spacings tend to change more mildly in response to the oxygen substitution. Among them, the change in $H_{\rm{Fe-As}}$ is of particular significance because it correlates directly with the orbital-dependent band structures, which could control the emergence of superconductivity.\cite{two dome theory}

In the following statement, we cite the nominal oxygen content rather than measuring the oxygen content directly. The reasons as follows: (1) The $a$-axis basically decreases upon oxygen doping linearly, which is in line with expectation as O$^{2-}$ is smaller than N$^{3-}$.\cite{radii} (2) No obvious impurity peak can be found in the XRD profile for $x\leq0.5$. (3) The validity of using nominal oxygen content is supported by the similar experiments in NdFeAsO$_{1-x}$F$_x$, which indicated that the measured fluorine content is basically identical to the nominal content within the solubility limit.\cite{Nd1111-F WDS} (4) Our XPS and EDX experiments showed that, when exposed to the atmosphere, the sample will adsorb oxygen which strongly interferes with the analysis of elemental ratios.\cite{ThFeAsN} As it is nearly impossible to isolate the sample from the air during the transfer, it is very difficult for us to accurately measure the oxygen content directly.

\subsection{Characterization of physical properties}

\begin{figure}
\includegraphics[width=8cm]{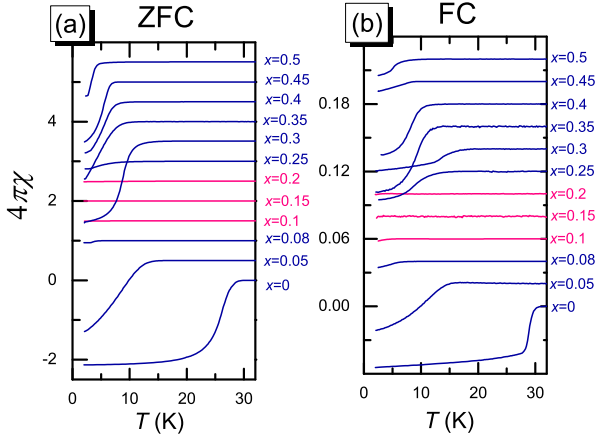}
\caption {(Color online)  (a) and (b): Magnetic susceptibility measured at 10 Oe near superconducting $T_\mathrm{c}$ for ThFeAsN$_{1-x}$O$_x$ samples. The $\chi-T$ curves of different samples are shifted successively along the vertical axis for comparison. The shift steps are 0.5 and 0.02 for zero-field-cooling (ZFC) and field-cooling (FC), respectively. }
\label{fig3}
\end{figure}

Figure \ref{fig3} (a) and (b) show the temperature dependence of magnetic susceptibility at $H=10$ Oe for ThFeAsN$_{1-x}$O$_x$ samples. As shown in the figure, the $T_\mathrm{c}$ value depends dramatically on the nominal oxygen content. To begin with, the oxygen substitution at only 5\% leads to the suppression of $T_\mathrm{c}^\mathrm{onset}$ from 30 K to 15.5 K. For the sample of $x=0.08$, $T_\mathrm{c}^\mathrm{onset}$ is further reduced to 6.7 K and the magnetic shielding fraction (MSF) at 2 K is merely about 5\%, indicating that most part of the sample is not superconducting. This is consistent with the transport property ($x=0.08$) in Figure \ref{fig5} which shows absence of zero resistance down to 2 K. In the doping range of $0.1\leq x\leq0.2$, no diamagnetic signal is observed above 2 K. When the nominal oxygen concentration reaches 25\%, superconductivity revives with $T_\mathrm{c}^\mathrm{onset}=13.5$ K. After that the $T_\mathrm{c}^\mathrm{onset}$ gets to a maximum of 17.5 K at $x=0.3$ (the large $\rm{MSF}\approx200\%$ is due to the demagnetization effect), and then gradually decreases to 7 K at $x=0.5$. For the last sample with $x=0.6$, superconductivity disappears again.

\begin{figure}
\includegraphics[width=8cm]{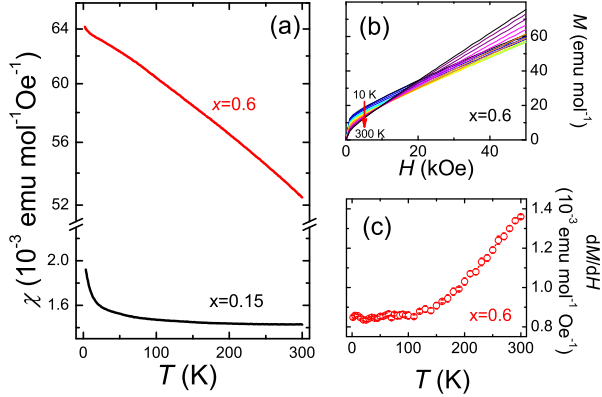}
\caption {(Color online) (a) Temperature dependence of magnetic susceptibility measured at 1 kOe for the samples of $x=0.15$ and $x=0.6$ respectively. (b) The isothermal magnetization curves for the sample of $x=0.6$ at various temperatures. (c) Derivative d$M$/d$H$ value as function of temperature. The d$M$/d$H$ values were obtained by a linear fitting for the high field (30 to 50 kOe) $M(H)$ data.}
\label{fig4}
\end{figure}

\begin{figure}
\includegraphics[width=7.5cm]{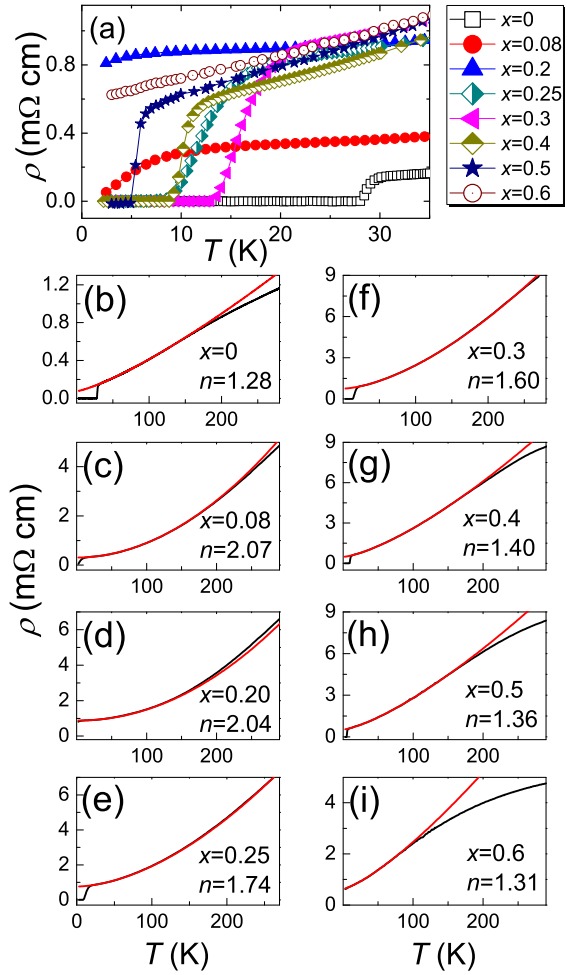}
\caption {(Color online) (a): Magnified $\rho-T$ curves near superconducting $T_\mathrm{c}$ for all the samples. (b)-(i): Temperature dependence of electrical resistivity for ThFeAsN$_{1-x}$O$_x$ samples. The red line fits the normal state resistivity using the equation $\rho(T)=\rho_0+AT^n$. }
\label{fig5}
\end{figure}

Figure \ref{fig4} shows the magnetic susceptibility measured at 1 kOe for the non-superconducting samples of $x=0.15$ and $x=0.6$. None of them can be fitted using Curie-Weiss Law. For $x=0.15$, the susceptibility value at 300 K is $1.42\times10^{-3}$ emu/mol. Upon cooling (above 150 K), the susceptibility is not very sensitive to the temperature, suggesting Pauli paramagnetic behavior (the low temperature upturn is ascribed to small amount of paramagnetic impurities). No anomaly indicating long-range magnetic ordering can be observed in the $\chi-T$ plot. For $x=0.6$, the room temperature susceptibility reaches $5.3\times10^{-2}$ emu/mol, which is significantly larger than that of LaFeAsO$_{1-x}$F$_x$ system ($\sim4\times10^{-4}$ emu/mol).\cite{susceptibility} To understand its origin, we performed isothermal magnetization measurements, which are shown in Figure \ref{fig4}(b). The plots of magnetization ($M$) versus applied field ($\mu_{0}H$) clearly indicate existence of ferromagnetic impurity, whose Curie point is higher than 300 K. This is not surprising, as the the XRD profile shows that the sample is not single-phase. Thus, the extrinsic ferromagnetic contribution to $\chi$ can be removed by using the derivative d$M$/d$H$ in the high-field regime instead of $\chi = M/H$. The d$M$/d$H$ values were obtained by a linear fitting for the high field (30 to 50 kOe) $M(H)$ data, which are shown in Figure \ref{fig4}(c). It can be seen that the d$M$/d$H$ value is $\sim$ 40-times smaller than the $\chi$ value at 300 K. The nearly linear d$M$/d$H$ above $\sim$160 K is consistent with that of ThFeAsN parent compound.\cite{ThFeAsN} Again, we cannot see any evident anomaly.

The temperature dependent resistivity near superconducting $T_\mathrm{c}$ is shown in Figure \ref{fig5}(a). The superconducting transition widths (defined as the temperature interval between 90\% to 10\% of normal-state resistivity) vary from 1 K to 5 K, which are widely observed in ``1111'' type polycrystalline samples.\cite{cxh, GdFeAsO-Th, LaFeAsO-H two dome, LaFeAsO-F two-dome} The broadening of the superconducting transition may be due to doping inhomogeneity.\cite{weak-link} To track the variation of superconductivity, here we pay attention to the temperature-dependent normal-state resistivity $\rho(T)$. Knowing that most of the iron-based superconductors exhibit strong anisotropy and the resistivity along the \emph{c}-axis could be as high as two orders of magnitude larger than that of the \emph{ab}-plane.\cite{Johnston review} So, when the current flows through a polycrystalline sample where the orientation of crystallites are supposed to be random with no preferred direction, most of the current will choose the path of low resistance route according to Kirchhoff's current law. Thus, the $\rho(T)$ behavior of a polycrystalline sample is mainly determined by the low-resistivity $\rho_{ab}$, which reflect the intrinsic property.\cite{cxh,anomaly-hosono} This is also supported by works in Ba$_{1-x}$K$_x$Fe$_2$As$_2$ system, where the resistivity behavior of a polycrystalline sample is identical to that of the \emph{ab}-plane in a single crystal.\cite{Ba122-K,ab-plane-2} Figure \ref{fig5}(b)-(i) shows the $\rho-T$ curves up to the room temperature. We fit the $\rho(T)$ data in the temperature range 20 K $\leq T\leq150$ K (for $x=0.6$, the fitting range changes to 20 K $\leq T\leq$ 100 K) using the equation $\rho(T)=\rho_0+AT^n$ where the exponent $n$ marks the scattering mechanism\cite{cxh,anomaly-hosono}. The $n$ value of the non-doped sample is 1.28, obviously deviated from 2.0 expected for a Landau-Fermi liquid. In the intermediate non-superconducting area (including $x=0.08$ with very small superconducting fraction), the $n$ value is close to 2.0. With the re-emergence of superconductivity, the exponent decreases rapidly to about 1.5. For the non-superconducting sample $x=0.6$, the exponent is 1.31.

\begin{figure}
\includegraphics[width=8cm]{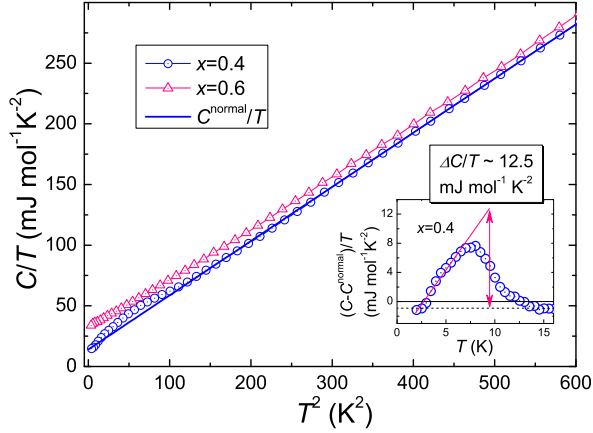}
\caption {(Color online) Temperature dependence of the specific heat in $C/T$ vs $T^2$ plot in the low-temperature region for samples of $x=0.4$ and $x=0.6$. The blue straight line fits the normal state specific heat using the equation $C^\mathrm{normal}(T)=\gamma T+\beta T^3$. The inset shows $(C-C^\mathrm{normal})/T$ vs. $T$ for $x=0.4$.}
\label{fig6}
\end{figure}

Figure \ref{fig6} show the temperature dependent specific heat for samples of $x=0.4$ and $x=0.6$. For the non-superconducting sample $x=0.6$, the $C/T$ vs $T^2$ plot slightly deviates from linearity below 5 K. Similar phenomenon was also observed in ThNiAsN superconductor, which is ascribed to Schottky anomaly of magnetic impurities and/or some nuclei.\cite{ThNiAsN} In the temperature range 5-25 K, the $C/T$ value is linearly related with $T^2$. This allows the estimation of Sommerfeld coefficient $\gamma=32.4$ mJ/mol/K$^2$, using the equation $C/T=\gamma+\beta T^2$. As a comparison, the $\gamma$ value in heavily electron-doped LaFe$_{0.5}$Co$_{0.5}$AsO system is only 1.68 mJ/mol/K$^2$, nearly 20 times smaller.\cite{LaFeAsO-Co Sefat} The enhancement cannot be solely ascribed to the change in density of state, which means that there is a significant increase in the effective mass of electrons. For the superconducting sample of $x=0.4$, the electronic specific heat in superconducting state can not be obtained by subtracting the specific heat measured in magnetic field, as the highest field available in our experiment (80 kGs) is not strong enough to suppress the superconducting state. Supposing that $C^\mathrm{normal}(T)=\gamma T+\beta T^3$, we fit the normal state specific heat ($C^\mathrm{normal}$) in the temperature range 11 K $\leq T\leq$ 25 K where the $C/T$ value is linearly related with $T^2$. The derived $\gamma$ is 14.1 mJ/mol/K$^2$. By subtracting $C^\mathrm{normal}(T)$ from the raw data, the specific heat jump due to the superconducting transition is revealed in the inset of Figure \ref{fig6}. One may note that, the entropy of the superconducting transition does not conserve up to $T_\mathrm{c}$. This suggest that the contribution from Schottky anomaly, which shows a broad peak at low temperatures,\cite{Schottky} also exist in the data.  Nevertheless, a jump of specific heat($\Delta C/T_\mathrm{c}=$12.5 mJ/mol/K$^2$) is evidently observed below 13 K. The thermodynamic transition temperature determined by an entropy-conserving construction is 9.2 K, conforming with the $T^\mathrm{zero}_\mathrm{c}$ in the Figure \ref{fig5}(a). Then, the $\Delta C/\gamma T_\mathrm{c}$ value is determined as 0.89, which is significantly smaller than the weak-coupling limit of BCS superconductors ($\Delta C/\gamma T_\mathrm{c} \approx 1.43$). The reduction of $\Delta C/\gamma T_\mathrm{c}$ value can be ascribed to an anisotropic superconducting gap within the $\alpha$-model of BCS theory.\cite{C-jump,C-jump-2} One may note an anisotropic superconducting gap was also proposed for the parent compounds ThFeAsN according to the measurements of specific heat($\Delta C/T_\mathrm{c}=$25 mJ/mol/K$^2$) as well as $\mu$SR spectroscopy.\cite{ThFeAsN-NMR,ThFeAsN-NMR-2} In the rigid-band picture, the hole band at the $\Gamma$ poind may gradually sink below the Fermi level upon electron doping, causing the shrink of the hole pocket.\cite{ThFeAsN-band1} This suggests that the anisotropic superconducting gap is more likely to be associated with the electron pocket.

\subsection{Doping phase diagram}

\begin{figure}
\includegraphics[width=8cm]{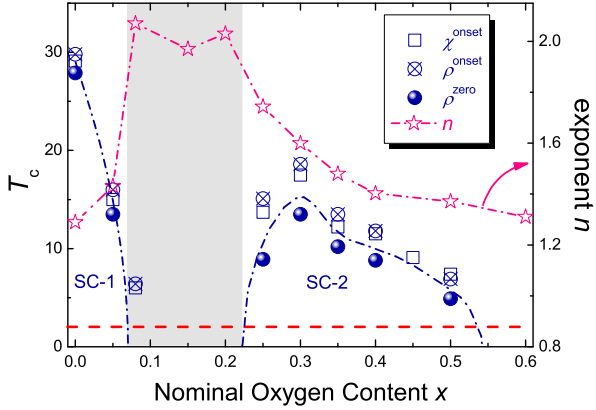}
\caption {(Color online) Superconducting phase diagram of ThFeAsN$_{1-x}$O$_x$ system. $\chi^\mathrm{onset}$, $\rho^\mathrm{onset}$ and $\rho^\mathrm{zero}$ represent the superconducting $T_\mathrm{c}$ determined by the onset of $\chi-T$ curve, the onset of $\rho-T$ curve and the zero-resistivity temperature respectively. The pink star (right axis) shows the exponent $n$ extracted from the normal state resistivity. The dashed line near the bottom axis indicates the lowest temperature available in our experiments.}
\label{fig7}
\end{figure}

We summarize superconducting $T_\mathrm{c}$ and the exponent $n$ in resistivity fitting as functions of the nominal oxygen content in Figure \ref{fig7}. The phase diagram shows two  superconducting areas separated by a non-superconducting zone with $0.1\leq x\leq0.2$. Previous heavily electron-doped cases show two superconducting domes that are mostly connected.\cite{LaFeAsO-H two dome,LaFeAsO-F two-dome,SmFeAsO-H} Isolated superconducting phases are shown in LaFeAs$_{1-x}$P$_x$O \cite{LaFeAsO-P two dome-1} in which the intermediate non-superconducting phase is magnetically ordered.\cite{LaFeAsO-P two dome-2} Another distinct feature of the present ThFeAsN$_{1-x}$O$_x$ system is that the maximum $T_\mathrm{c}$ for the second superconducting region is significantly lowered, in sharp contrast with other systems mentioned above which shows the opposite. Interestingly, the emergence of superconductivity seems to correlate with the exponent $n$, which tends to deviate from 2.0 when superconductivity emerges. Nevertheless, although the $n$ value of the sample of $x$ = 0.6 is 1.31, it does not superconduct.

The position of the non-superconducting phase in the phase diagram is another interesting issue, which could be related to the crystal structure of FeAs layers. Matsuishi \textit{et al}.\cite{SmFeAsO-H} show that the $T_\mathrm{c}$ valley between the two superconducting domes in LaFeAsO$_{1-x}$H$_x$ and SmFeAs$_{1-y}$P$_y$O$_{1-x}$H$_x$ systems locate at the electron doping level of $x_{\mathrm{val}}\approx0.16$ in which the diagonal As$-$Fe$-$As bond angle is $\alpha\approx113^\circ$. Interestingly, the correlation ($x_{\mathrm{val}}$, $\alpha$) is independent of both phosphorus doping and lanthanide species.\cite{SmFeAsO-H} Coincidentally, the non-superconducting area in ThFeAsN$_{1-x}$O$_x$ is centered at $x=0.15$ with the same bond angle of $\alpha=113^\circ$. Nevertheless, their ``starting points'' are quite different: the non-doped parent phases of $Ln$FeAsO show an AFM ground state, while no magnetic order is detected in ThFeAsN.\cite{ThFeAsN-Mossbauer,ThFeAsN-neutron,ThFeAsN-NMR} This implies that, aside from the electron doping, there are additional factors controlling the ground states of ThFeAsN$_{1-x}$O$_x$.

We propose that the built-in chemical pressure could play an important role. Firstly, the axial ratio of ThFeAsN ($c/a\approx2.11$) is the lowest among 1111-type iron arsenides.\cite{review-1111} Secondly, the $H_{\rm{Fe-As}}$ value (1.305 {\AA}) is the smallest and, the $\alpha$ value (114.2$^\circ$) is the largest, among FeAs-layer based compounds.\cite{review-Ganguli} All these suggest that ThFeAsN bears an internal chemical pressure exerted along the $c$-axis. This viewpoint is supported by the recent structural analysis which shows that the $H_{\rm{Fe-As}}$ value in ``1111'' phases tends to decrease linearly with increasing physical pressure.\cite{La1111H-pressure} So, the parameter $H_{\rm{Fe-As}}$ can be used as an indicator of the uniaxial chemical pressure, and the increase in $H_{\rm{Fe-As}}$ upon oxygen doping (Fig. 2(a)) suggests that the chemical pressure is gradually released in ThFeAsN$_{1-x}$O$_x$. This partly explains the rapid suppression of superconductivity at low doping (note that the oxygen substitution also introduces electrons, therefore the AFM ground state cannot be recovered). Thus, the position and the width of the non-superconducting window on ThFeAsN$_{1-x}$O$_x$ phase diagram are actually determined by a combined effect of chemical pressure and electron doping. The lowered maximum $T_\mathrm{c}$ of the second superconducting window could be caused by the remaining uniaxial chemical pressure.

Now let us discuss the origin of the two superconducting area in ``1111'' systems. For the first (left) superconducting region, either carrier doping\cite{Hosono LaFeAsO-F} or ``applying" chemical pressure with isovalent P/As doping\cite{wcao} suppresses the magnetic order in the parent compounds, after which superconductivity appears with remaining spin fluctuations in the normal state.\cite{fluctuation1,fluctuation2,fluctuation3,fluctuation4} This naturally leads to the picture that superconductivity is due to spin fluctuations associated with the nearby magnetic phase.\cite{F-picture1,F-picture2,F-picture3} For the second (right) superconducting area, the relationship between superconductivity and magnetism becomes ambiguous. On the one hand, recent studies on LaFeAsO$_{1-x}$H$_x$ found a second AFM state below 100 K on the right side of the second superconducting dome, accompanied by an orthorhombic lattice distortion due to As atom off-centering.\cite{AFM2,AFM3} In LaFeAs$_{1-x}$P$_x$O, a second AFM state was also observed, but it locates at about half doping, in between the two superconducting domes.\cite{LaFeAsO-P two dome-2} Both cases still suggest a spin-fluctuation scenario for the second superconducting dome.\cite{LaFeAsO-P two dome-2,AFM2,AFM3} However, on the other hand, the study on heavily doped LaFeAsO$_{1-x}$F$_x$ found neither AFM ordering nor low-energy magnetic fluctuations.\cite{LaFeAsO-F two-dome} Accordingly, an orbital-fluctuation mechanism was proposed for the second superconducting dome.\cite{LaFeAsO-F two-dome, two dome theory} As for ThFeAsN$_{1-x}$O$_x$, a long-range magnetic order seems unlikely according to the susceptibility measurement. In this aspect, further investigations using NMR, $\mu$SR or neutron diffractions are highly needed.

\section{Conclusion}

To summarize, we have successfully realized heavily electron doping in ThFeAsN$_{1-x}$O$_{x}$ ($0\leq x\leq0.6$) without using high-pressure synthesis. Our resistivity and susceptibility measurements reveal a peculiar phase diagram showing two isolated superconducting regions with maximum $T_\mathrm{c}^\mathrm{onset}$ of 30 K at $x=0$ and 17.5 K at $x=0.3$, respectively. The absence of superconductivity in $0.1\leq x\leq0.2$ basically coincides with the structural anomaly featured with a minimum of $c$-axis as well as a kink in the As height. We argue that the built-in uniaxial chemical pressure, in addition to the electron doping, plays an important role for the quenching of superconductivity within $0.1\leq x\leq0.2$ as well as for the lowered $T_\mathrm{c}^{\mathrm{max}}$ of the second superconducting area. The specific-heat jump for the sample $x=0.4$ implies anisotropic superconducting gap for the electron Fermi pocket. The present system supplies additional information for describing the global electronic phase diagram in iron-based superconductors, which could help to ultimately understand the superconducting mechanism.

\ack

This work was supported by NSF of China (Grant Nos. 11304183 and 11190023), the National Key Research and Development of China (No. 2016YFA0300202), the Project of Shandong Province Higher Educational Science and Technology Program (No.J17KA183), and the open program from Wuhan National High Magnetic Field Center (2016KF03).

\end{document}